\documentclass[aps,twocolumn,showpacs,groupedaddress,amssymb]{revtex4}
\usepackage{bm}
\usepackage{amsmath}

\begin{document}

\title[Short Title]{Electronic States of Boron in Superconducting MgB$_2$ 
Studied by $^{11}$B NMR}
\author{A.~Gerashenko, K.~Mikhalev, S.~Verkhovskii}
\affiliation{Institute of Metal Physics, UB RAS, 620219
Ekaterinburg GSP-170, Russia}
\author{T.~D'yachkova, A.~Tyutyunnik, V.~Zubkov}
\affiliation{Institute of Solid State Chemistry, UB RAS,
620219 Ekaterinburg, Russia
}
\date{\today}

\begin{abstract}
NMR spectra and nuclear spin-lattice relaxation rate $T_1^{-1}$
of $^{11}$B have been measured in superconducting polycrystalline
MgB$_{2}$ with $T_c^{ons}\cong39.5K$. It is shown that $(T_{1}T)^{-1}$
and the Knight shift $K_{s}$ are independent of temperature and
nearly isotropic above $T_c$. Both of these quantities
are decreased gradually in going to superconducting state. According
NMR data the density of states (DOS) near the Fermi level is
flat at the scale $\approx$ 500K. Some conclusions on the orbital
content of the DOS at the Fermi level was done and compared with
the results of the band structure calculations.
\end{abstract}
\maketitle

Discovery of superconductivity in magnesium diboride MgB$_2$ showing
remarkable high
temperature of superconducting transition
($T_c\cong$ 40K)~\cite{Akimitsu} stimulated intensive studies
of the electron properties of this new medium-$T_c$ superconductor.
Crystal structure
of MgB$_2$ (space group symmetry $D_{6h}^1$) is well known and relates
to the hexagonal AlB$_2$-type~\cite{Goldschmidt}.
Boron forms a primitive honeycomb lattice and
pronounced $B-B$ covalent bonding  creates graphite-like sheets of
boron separated
by hexagonal layer of Mg. Calculations of electronic band structure
show that the
bands near the Fermi level are derived mainly from $2p_{x,y}$ bonding
orbitals of boron~\cite{Ivanovskii,Kortus}.
As predicted the dispersion of these bands is extremely small near
the $\Gamma$ point of the
Brillouin zone. They form two small cylindrical Fermi surfaces around
$\Gamma$-A line.
Due to their 2D character they contribute more than one third of the total
DOS - $N(0)_{E_F}$=0.199 (eV at.B)$^{-1}$. Sizeable electron phonon
coupling is predicted for
electrons in these $2p_{x,y}$ bands~\cite{Kortus} and strong boron
isotope effect reported in~\cite{Bud'ko} is
in favor that MgB$_2$ being phonon mediated superconductor.
Electron-phonon coupling
constant is proportional to DOS at the Fermi level and it is quite
important to have
experimental estimations of this quantity in MgB$_2$.

We have measured NMR line shift $(K)$ and nuclear spin-lattice relaxation
rate $(T_1^{-1})$ of $^{11}$B. For boron in metallic MgB$_2$ both these
quantities
contain the contributions determined by hyperfine coupling with
conducting electrons of s and p bands. The hyperfine fields of
s-electrons leads to isotropic contributions to the Knight shift $(K_s)$
and $(T_1^{-1})$. While hyperfine coupling with conducting electrons
of p bands
leads to anisotropic contribution  both to the NMR line shift $(K_{ax})$
and
the nuclear spin-lattice relaxation rate of $^{11}$B.
This anisotropic contribution to
$(T_1^{-1})$ depends on the orbital content of $2p_i$ electrons,
having near the
Fermi energy appropriate DOS $N_i=f_i*N(E_F)$ \cite{Abragam,Obata}.

Polycrystalline MgB$_2$ was prepared using a mixture of magnesium chips
and amorphous boron pressed into pellets and placed in a tantalum crucible.
The pellets were subjected to a two step annealing in purified helium flow
$(\lg Po_2=-19)$. A first step pellets were annealed at $T_{an}$=800C
for 1 hour and
resulting phase content was the following: MgB$_2$ (78.6\%); Mg(16.6\%);
MgO (4.7\%)
BN (traces). After succeeding annealing at $T_{an}$=875C for 1 hour
(evaporation of magnesium) the final phase content was obtained:
MgB2 (89.9\%);
MgO (5.2\%); BN (1.7\%). Quantitative phase analysis of the samples
was carried out
using X-ray powder diffraction data collected with a STADI-P
diffractometer
(STOE, Germany) and PDF2 data base (ICDD, USA). AC magnetization
measurements show
the SC transition with $T_C^{ons} = 39.5$K and $\Delta T_c=2.5$K.

NMR measurements were carried out on a pulse spectrometer over the
temperature
range of 5-300K in magnetic fields $H_0=2.1$ and 9.1~T.
The spectra were obtained by
Fourier transformation of the second half of the spin-echo signal
followed the
$(\pi /2)_x- \tau _{del} -(\pi)_x$ pulse sequence. The broad spectra
exceeding the frequency band
excited by rf-pulse were measured by summation of an array of
Fourier-signals of an
echo accumulated at different equidistant operating frequencies.
The components of
magnetic shift tensor $(K_{iso},K_{ax})$ for $^{11}$B (I=3/2) as well
as the electric field gradient
(EFG) parameters - quadrupole frequency $\nu_Q$ and asymmetry
parameters $\eta$- were determined
by computer simulation of the measured NMR spectra. The powder pattern
simulation program takes into account the quadrupole coupling
corrections up to
the second order of the perturbation theory. Positions of the features
of both
central line (transition $m=-\frac12 \leftrightarrow +\frac12$) and
satellite lines
(transitions $m=\pm \frac12 \leftrightarrow \pm \frac32$) were involved
into consideration.
The diamond form of BN was used as a reference.

The boron atoms in MgB$_2$ unit cell are located at the sites
with the axial symmetry of the nearest environment. The $^{11}$B
nuclear has a spin $^{11}$I=3/2 and electric quadrupole moment
$^{11}Q=0.04065 \cdot 10^{-24}$ cm$^2$~\cite{Landolt}.
Thus the resonance frequency of $^{11}$B NMR-probe is determined
by both hyperfine magnetic interaction and the interaction of
quadrupole moment $^{11}$Q with electric field gradient (EFG) V$_{zz}$,
created at the nuclear site by the electronic and ionic environments.

The NMR spectrum of $^{11}$B including all transitions is shown in
Fig. 1 for $T=300$K. Simulation of the powder patterns NMR line
shape allowed us to determine with a reasonable accuracy spherical
components of the magnetic shift ($K_{iso}$, $K_{ax}$) and the EFG
$(\nu _Q, \eta)$
tensors: $K_{iso}$=175(15)~ppm, $K_{ax}<30$~ppm;
$\nu_Q=\frac{3eQ}{2I(2I-1)h}V_{zz}$ =
0.828(10)MHz; $\eta$=0. The measurements at different $T$
have demonstrated that magnetic shift and EFG parameters are
independent of temperature in normal state down to $T_c(H)$
(see Fig. 2). Similar observation for magnetic shift of the $^{11}$B
NMR central line was reported recently in Ref.~\cite{Kotegawa}.

Magnetic shift is decreased gradually in going to superconducting
state and at $T=20$K${\approx}T_c$/2 its magnitude becomes
close to zero. Unfortunately we can not evaluate precisely the
magnitude of the Knight shift from the data obtained, since additional
diamagnetic contribution due to supercurrents arisen around vortexes
should be involved into consideration. At present time one may
conclude that $K_{s}$(20K) is less than 100~ppm.

Nuclear spin-lattice relaxation rate of $^{11}$B was measured
using the inversion-recovery technique and working at the
frequency domain. Amplitude of rf magnetic field in the pulse
exceeded 100 Oe. The measurements were performed at different
point of spectrum to study anisotropy of $T_1(\theta)$. Here
$\theta$ is the angle between $\bm{c}$-axis and
$\bm{H}_{\bm{0}}$. The measurements at the peak of central
line give the angle-averaged magnitude of $T_1$. In this case
$T_{1}$ was determined by fitting $m(t)$ data to the following
expression~\cite {Narath}:
\begin{eqnarray}
\begin{array}{rcl}
m(t)&=&\frac{M(\infty)-M(t)}{M(\infty)} \\
    &=&\frac{1}{10}\exp\left(-\frac{t}{T_1}\right) +\frac{9}{10}\exp\left(-\frac{6t}{T_1}\right).
\end{array}
\end{eqnarray}
In measurements at a peak ($\bm{c}\perp\bm{H}_{\bm{0}}$) and a step
($\bm{c}\parallel\bm{H}_{\bm{0}}$) of the high-frequency satellite powder
pattern (see inset in Fig.3) the magnitude of $T_1$ was determined
by fitting $M(t)$ data to the another expression~\cite {Narath}:
\begin{eqnarray}
\begin{array}{rcl}
m(t)&=&\frac{M(\infty)-M(t)}{M(\infty)}=\frac{1}{10}\exp\left(-\frac{t}{T_1}\right) \\
    &+&\frac{5}{10}\exp\left(-\frac{3t}{T_1}\right)
       +\frac{4}{10}\exp\left(-\frac{6t}{T_1}\right).
\end{array}
\end{eqnarray}
The temperature dependence of the product ($T_1T)^{-1}$ measured
at the central line is shown in Fig. 4. In the normal state the
magnitude of ($T_1T$)=155(5)sK is found as independent of temperature.
A slightly larger value of ($T_1T$) was reported in
Ref.~\cite{Kotegawa} for bulk
sample of MgB$_2$. Independent of temperature product ($T_1T)^{-1}$
may be considered as an evidence for a flatness of the $N(E)$ curve
near the Fermi energy at the scale of $\sim$500K.

A rather unexpected result is obtained in measurements of $T_1$
at the crystallites which $\bm{c}$-axis is differently oriented
with respect $\bm{H}_{\bm{0}}$. As seen at Fig. 3 the $M(t)$ data being
measured at differently oriented crystallites are fitted well
to the expression (2) with the very same adjustable parameter
$T_{1}$. Corresponding magnitude of the product ($T_1T$)=155(5)sK as
was obtained for central line. Using the results of band calculations
we expect to find the ratio
$T_1(c\perp H_0)/T_1(c\parallel H_0)\approx$ 1.2,
if to assume that nuclear spin lattice relaxation rate is monitored
exclusively by hyperfine dipolar and orbital interaction between
nuclear magnetic moment and p-electrons of the partly filled
$p_{x,y}$ bands. The results obtained means that $T_1^{-1}$ is determined
by the Fermi contact interaction with conducting electrons of
s-band or both of three partial $p_{i}$ DOS are equal
(f$_{x}$=f$_{y}$=f$_{z}$).

At temperature below $T_c$  nuclear spin-lattice relaxation
rate deviates of the Korringa behavior. No evidences for the
Hebel -- Slichter~\cite{Hebel} coherence peak was seen in measurements
near $T_c$
at external field of 2.1T. At temperature below 20K we approximated
$T_{1}$ data by expression in the thermally-activation exponential
form and found the ratio $2\Delta /k_BT_c>2.5$ assuming
isotropic magnitude of the superconducting energy gap in MgB$_{2}$.

The authors benefited much by suggestions made by Prof.B.N.~Goshchitskii
and acknowledge Dr.N.I.~Medvedeva and Dr.A.L.~Ivanovsky for helpful
discussions. Work is supported by the Russian State contract
No107-1(00)-P (order No22) and Russian State Program of Support
of Leading Scientific Schools (project No 00-15-96581).

\begin{figure}
\caption{$^{11}$B NMR spectrum of MgB$_2$ measured in the magnetic field
of 9.123T at room temperature. Simulated powder pattern of the
spectrum is shown by solid line drawn in red colour.}
\label{}
\end{figure}

\begin{figure}
\caption{The temperature dependence of $^{11}K_{iso}$ measured at the
external fields of 9.123T ($\circ$) and 2.113T ($\bullet$).
The value of $T_c$ in the field is pointed out by an arrow.}
\end{figure}

\begin{figure}
\caption{Recovery of $^{11}$B nuclear magnetization in measurements
of $T_1$  using the inversion-recovery pulse sequence.
Irradiated regions of the $^{11}$B spectrum related to the
crystallites which c-axis is differently oriented with respect
to $\bm{H_0}$ are shown in inset. }
\end{figure}

\begin{figure}
\caption{The temperature dependence of $(^{11}T_1T)^{-1}$ product vs $T$
in MgB$_2$
measured at the external magnetic fields of 9.123T ($\circ$) and
2.113T ($\bullet$). The dotted straight line is a guide for the eyes
to show that the product $T_1T$ is independent of $T$.}
\end{figure}

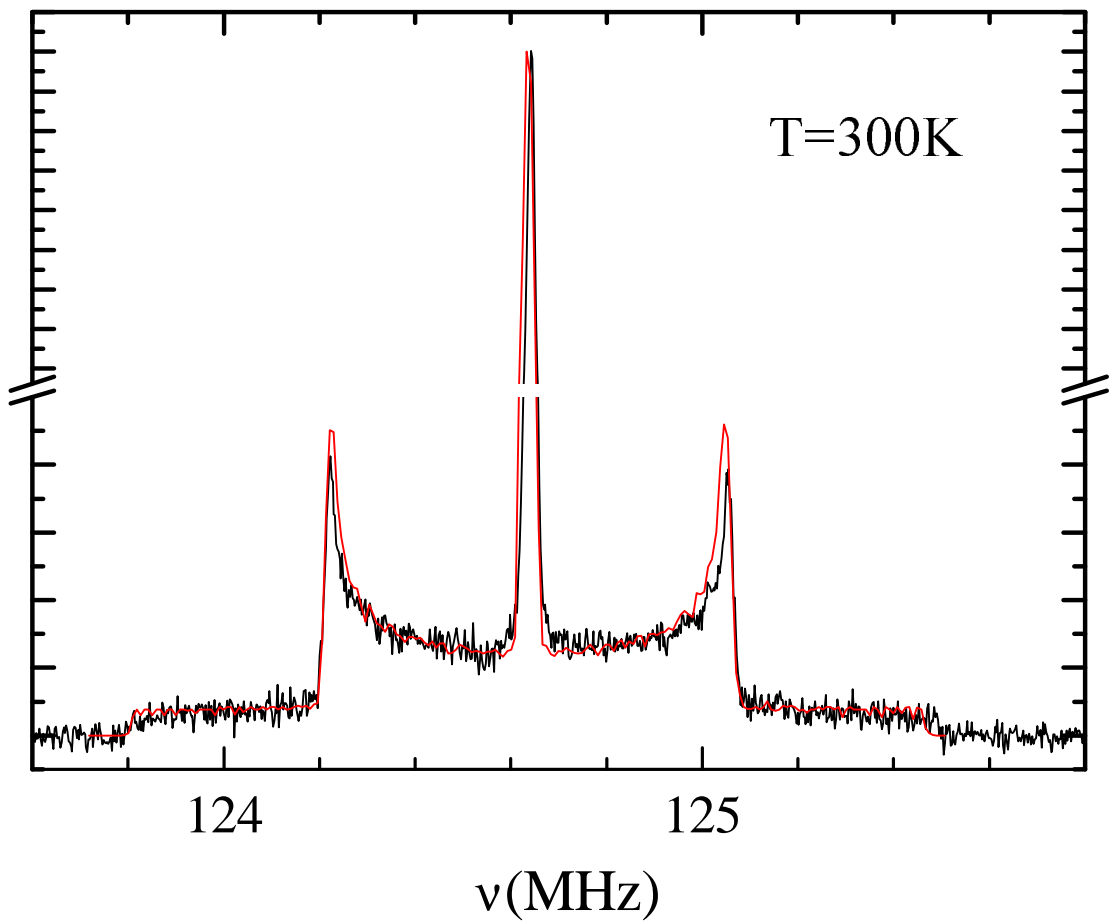
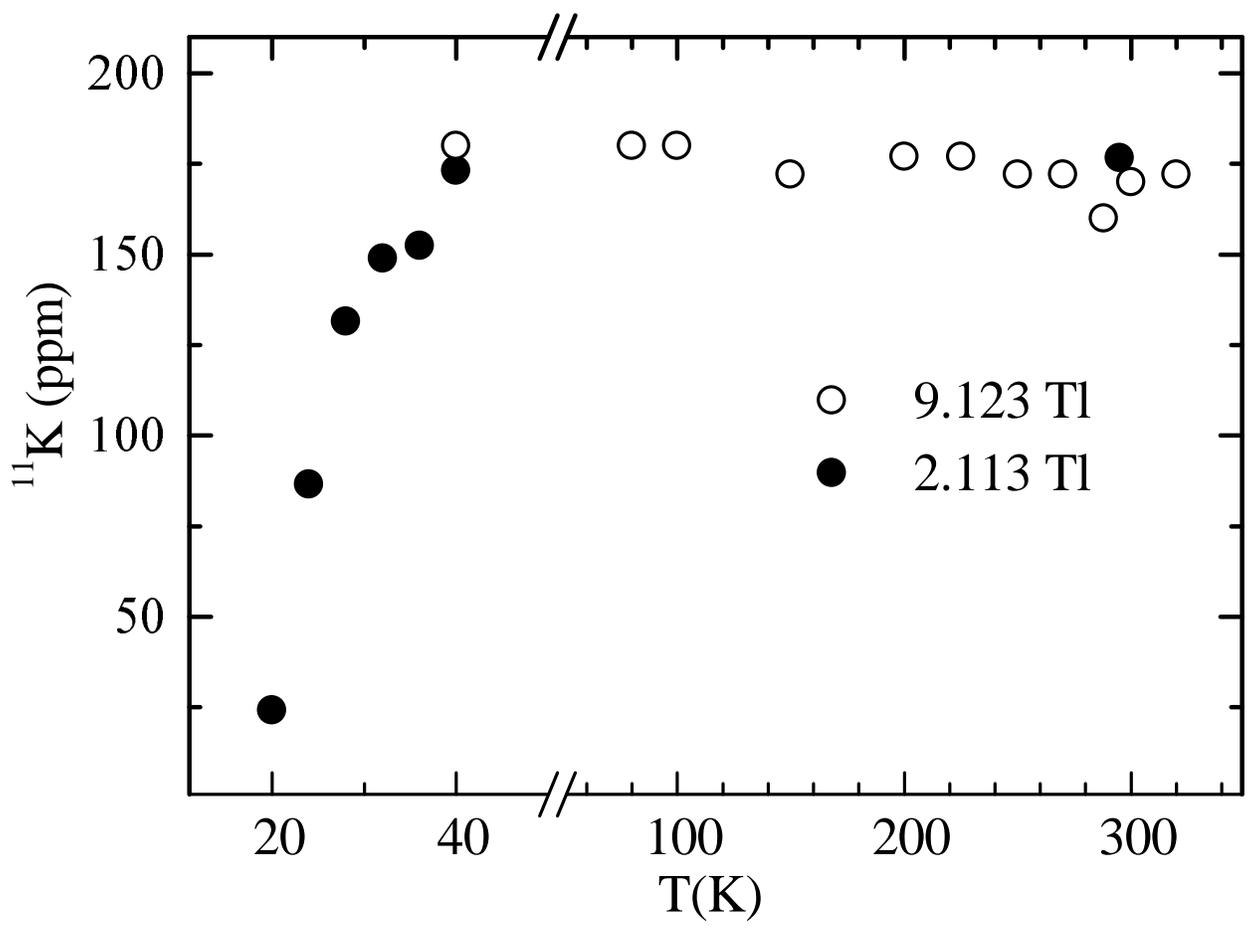
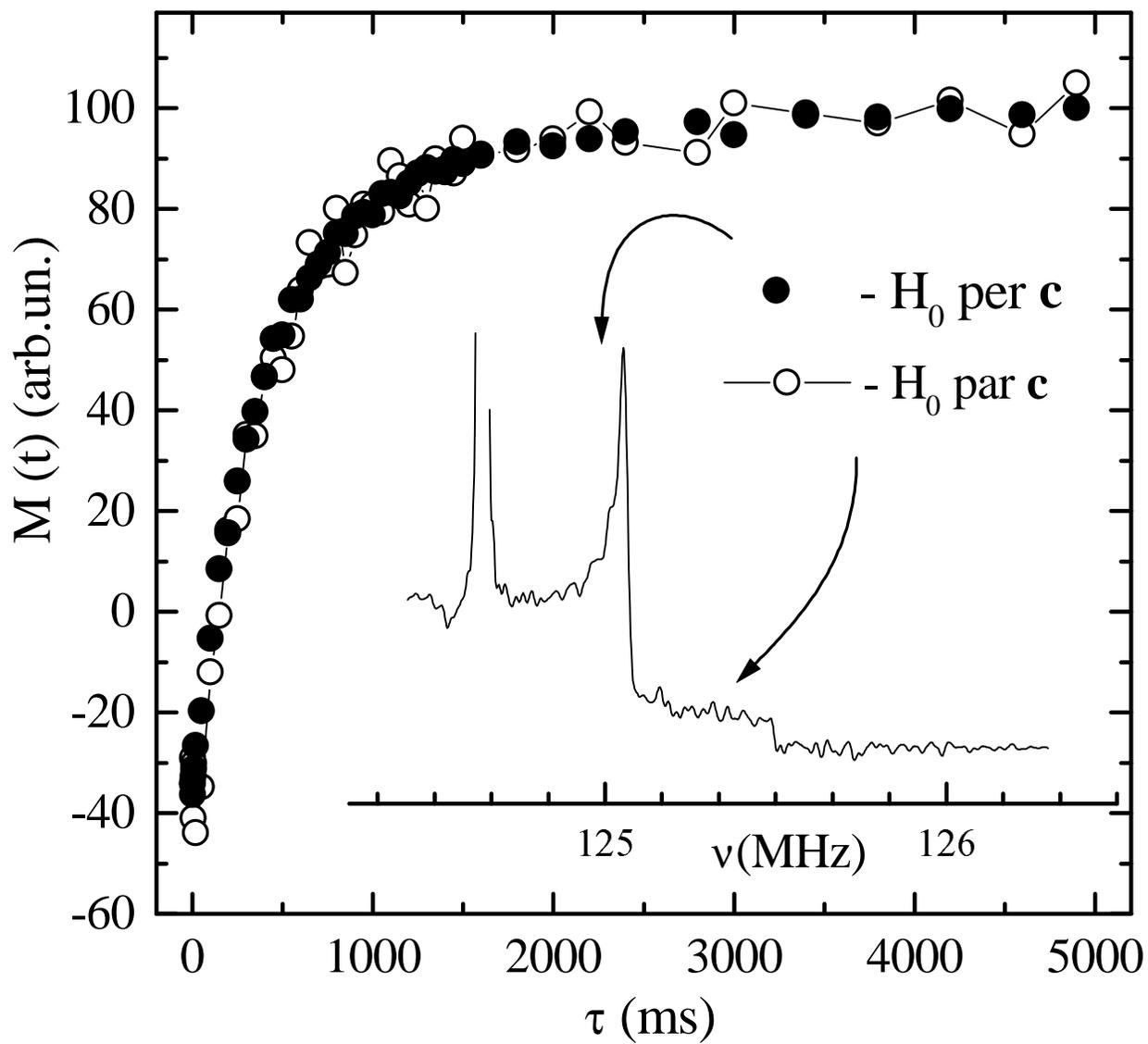
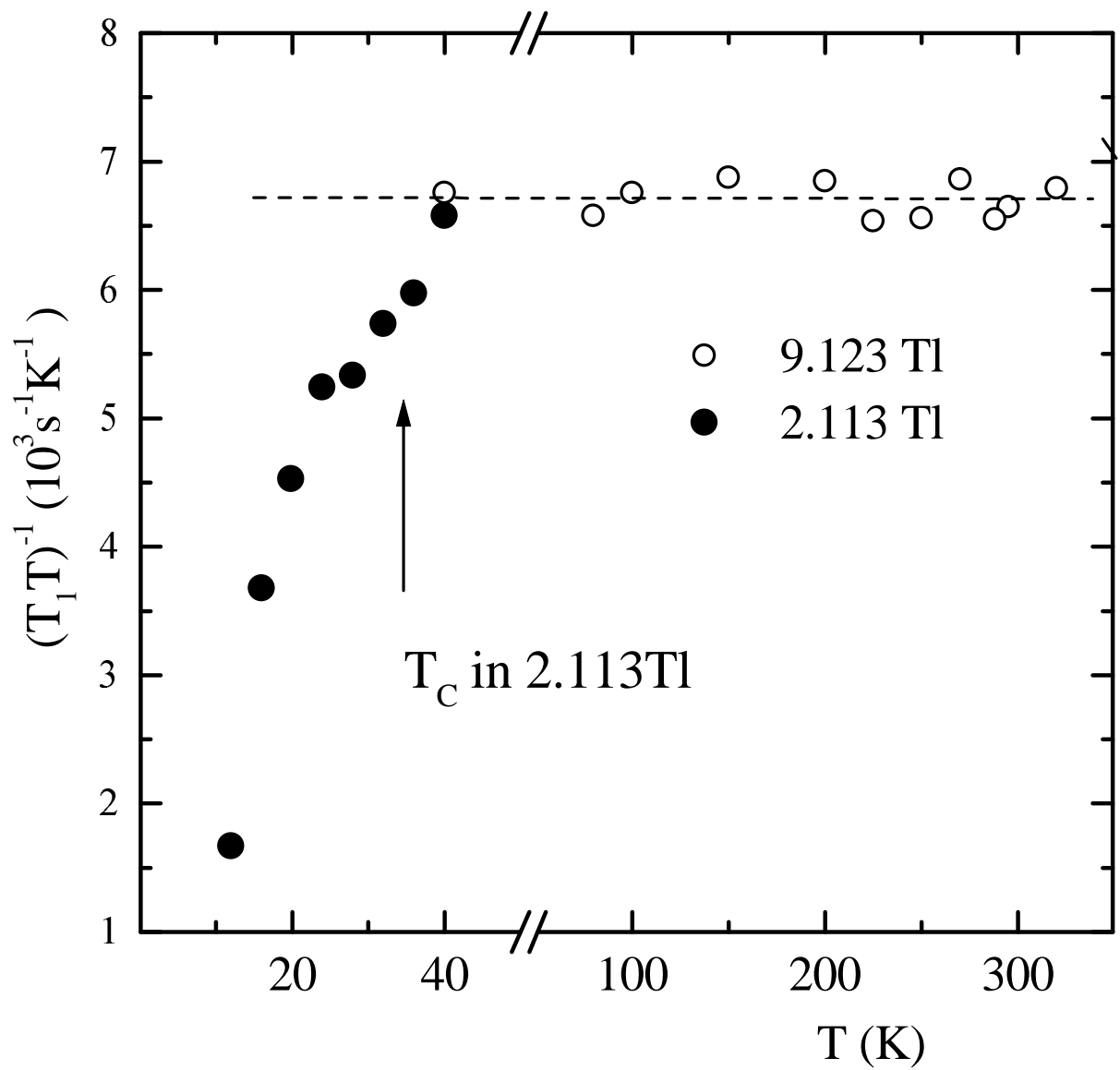
\end{document}